\documentclass{article}
\usepackage{spconf,amsmath,graphicx}

\usepackage{color}
\usepackage{enumitem}
\usepackage{hyperref}
\usepackage{multirow}
\usepackage{etoolbox,siunitx}
\robustify\bfseries
\usepackage{booktabs}
\usepackage{balance}
\usepackage{amssymb}
\usepackage{bm}
\usepackage{dsfont}
\usepackage{arydshln}
\usepackage{caption}
\usepackage{cite}

\let\OLDthebibliography\thebibliography
\renewcommand\thebibliography[1]{
  \OLDthebibliography{#1}
  \setlength{\parskip}{0pt}
  \setlength{\itemsep}{0pt plus 0.3ex}
}
\def\RR{{\mathbb R}}
\title{Hyperbolic Audio Source Separation}
\name{Darius Petermann$^{1,2}$, Gordon Wichern$^1$, Aswin Subramanian$^1$, Jonathan Le Roux$^1$\thanks{This work was performed while D.~Petermann was an intern at MERL.}}
\address{$^1$Mitsubishi Electric Research Laboratories (MERL), Cambridge, MA, USA\\
$^{2}$Indiana University, Department of Intelligent Systems Engineering, Bloomington, IN, USA
}

\begin{document}
\ninept
\maketitle
\setlength{\abovedisplayskip}{2.5pt}
\setlength{\belowdisplayskip}{2.5pt}

\begin{abstract}
We introduce a framework for audio source separation using embeddings on a hyperbolic manifold that compactly represent the hierarchical relationship between sound sources and time-frequency features. Inspired by recent successes modeling hierarchical relationships in text and images with hyperbolic embeddings, our algorithm obtains a hyperbolic embedding for each time-frequency bin of a mixture signal and estimates masks using hyperbolic softmax layers. On a synthetic dataset containing mixtures of multiple people talking and musical instruments playing, our hyperbolic model performed comparably to a Euclidean baseline in terms of source to distortion ratio, with stronger performance at low embedding dimensions. Furthermore, we find that time-frequency regions containing multiple overlapping sources are embedded towards the center (i.e., the most uncertain region) of the hyperbolic space, and we can use this certainty estimate to efficiently trade-off between artifact introduction and interference reduction when isolating individual sounds.
\end{abstract}
\begin{keywords}
audio source separation, hyperbolic space, speech, music, sound hierarchy
\end{keywords}
\section{Introduction}
\label{sec:intro}

A fundamental paradigm in deep learning-based audio source separation algorithms is the idea of applying a mask to a feature representation (e.g., a magnitude spectrogram or learned basis) of an audio mixture signal~\cite{erdogan2015mask, hershey2016dpcl, luo2019conv, wang2018supervised}. By inverting or decoding the masked feature representation, we obtain the isolated time-domain source signals.  While techniques that learn feature encoders and decoders directly based on waveform signals have achieved impressive performance~\cite{luo2019conv, defossez2021hybrid}, they lack interpretability compared to techniques based on time-frequency (T-F) representations such as the short-time Fourier transform (STFT) spectrogram~\cite{heitkaemper2020demystifying, parikh2022harmonicity}. %
Among these,
algorithms such as deep clustering~\cite{hershey2016dpcl} and deep attractor networks~\cite{luo2018dan} learn an embedding vector for each T-F bin, and create masks using classifiers and/or clustering algorithms to assign embeddings to sources. A fundamental problem for these approaches then becomes how to best learn a discriminative embedding for each T-F bin.

\begin{figure}
    \centering
        \includegraphics[width=1.0\linewidth]{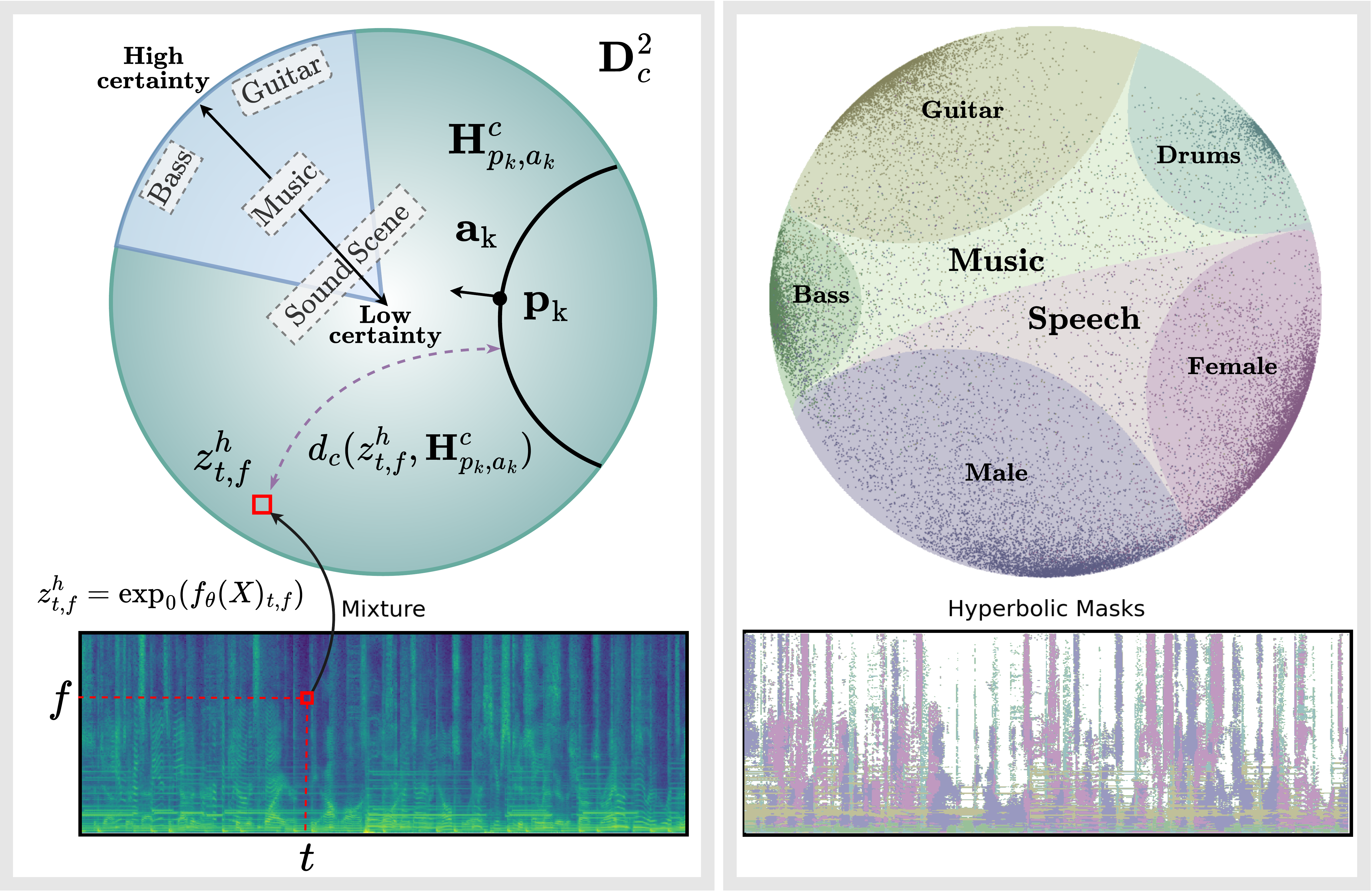}\vspace{-.1cm}
    \caption{Illustration of hyperbolic source separation. Left plot demonstrates the process of taking a T-F bin from a mixture spectrogram $X_{t,f}$ and projecting its Euclidean embedding to $z^h_{t,f}$ on the Poincar\'e ball (red squares); the dashed purple arrow represents the distance of the resulting embedding to the hyperplane $\text{H}^c_{p_k,a_k}$. Right plot shows the Poincar\'e ball with the resulting decision boundaries from a trained model (top). The mixture embeddings (scatters) are colored-coded according to the maximum softmax layer output. The predicted T-F masks for each of the leaf sources (bottom) are plotted  similarly, and silent bins (-40 dB and below) are colored in white. }\vspace{-.6cm}
    \label{fig:overall_example}
\end{figure}

In this work, we take inspiration from recent advances in modeling language~\cite{suzuki2021ordinal, shimizu2021hyperbolic, sun2021hgcf}, graphs~\cite{liu2019hyperbolicGN,bachmann2020constantCG,sun2021hgcf}, and images~\cite{Khrulkov2020CVPR,chami2020tree,guo2021free, Atigh_2022_CVPR,weng2021longtail, suris2021learning} in hyperbolic space, and explore their relevance for audio source separation. Unlike Euclidean spaces, hyperbolic spaces have an inherent ability to infer hierarchical representations from data with very little distortion~\cite{salar2018representations,sarkar2011distortion}. Hierarchical and tree-like structures are ubiquitous in many types of audio processing problems such as musical instrument recognition~\cite{essid2006hierarchical, garcia2021leveraging} and separation~\cite{manilow2020hierarchical}, speaker identification~\cite{jieun2022speechpoincare}, and sound event detection~\cite{jati2019hierarchy, cramer2020chirping}. However, all of these approaches model the hierarchical information globally by computing a single embedding vector for an entire audio clip. Recent work in image segmentation~\cite{Atigh_2022_CVPR}  learns a hyperbolic embedding at the pixel level, and we take a similar approach by computing a hyperbolic embedding for each T-F bin of an audio mixture %
spectrogram, as illustrated in Fig.~\ref{fig:overall_example}.

As has been shown by many recent computer vision studies~\cite{Khrulkov2020CVPR, suris2021learning}, hyperbolic network layers~\cite{ganea2018hyperbolicnn} can be added to the end of existing network architectures to learn and classify hyperbolic embeddings. In addition to benefits in terms of hierarchical modeling, hyperbolic layers can learn strong embeddings at low dimensions~\cite{jieun2022speechpoincare}, represent uncertainty~\cite{Khrulkov2020CVPR, suris2021learning}, and estimate features such as edges in images based on their distance to the origin in the hyperbolic space~\cite{Atigh_2022_CVPR}. Our main contribution is to explore whether similar observations are applicable to hyperbolic embeddings of T-F bins, and we propose new avenues for interpreting and interacting with hyperbolic embeddings in audio source separation applications.

Specifically, we use a simulated dataset containing multiple overlapping speakers and musical instruments to illustrate our approach as shown in Fig.~\ref{fig:overall_example}. We classify each T-F bin embedding using hyperbolic softmax layers~\cite{ganea2018hyperbolicnn}, and each T-F bin is further assigned to both a parent and child class using a hierarchical softmax~\cite{Atigh_2022_CVPR}. We compare various widely used source separation loss functions, but find that energy-weighted cross-entropy performs best. Corroborating work in application domains such as computer vision and speaker identification, we find that performance with low embedding dimensions is relatively strong in hyperbolic space. In terms of uncertainty, we show how the distance to the origin can be used to approximate computationally expensive certainty estimates obtained with Monte-Carlo dropout~\cite{Mukhoti2018EvaluatingBD} using only a single forward pass. We also observe that T-F bins containing multiple overlapping sources are consistently embedded near the origin (i.e., they are considered the most uncertain embeddings). We then provide experiments demonstrating how this uncertainty can be exploited to navigate the trade-off between interference reduction and artifact introduction, which is fundamental to audio source separation.

\vspace{-.1cm}
\section{Hyperbolic Audio Source Separation}
\label{sec:hss}
\subsection{Embedding-based source separation}

We consider the problem of separating a mixture audio signal into $K$ sources of interest. While we later focus more precisely on class-based separation, where the sources belong to $K$ different sound classes, we first give a general presentation of separation approaches relying on time-frequency (T-F) embeddings.

Source separation is often formulated as a mask-inference problem \cite{erdogan2015mask} in a time-frequency domain, such as that obtained by a short-time Fourier transform (STFT), where one seeks to obtain the complex spectrogram $S^k \in \mathbb{C}^{T \times F}$ of each of the $K$ sources by multiplying element-wise the spectrogram of the mixture $X\in \mathbb{C}^{T \times F}$ with a real-valued mask $M^k \in \mathbb{R}^{T \times F}$: $\hat{S}^k = M^k \odot X.$
A common way to obtain the mask $M^k_{t,f}$ at time-frequency bin $(t,f)$ relies on first computing an $L$-dimensional Euclidean embedding $z^{\text{e}}_{t,f}\in \RR^L$ from $X$ using a deep neural network $f_{\theta}(.)$ parameterized by $\theta$, such that $Z^{\text{e}}=f_{\theta}(X)\in \mathbb{R}^{T \times F \times L}$.
These embeddings can be projected to dimension $K$ and passed through mask-output non-linearities such as sigmoid or softmax, or clustered in $\RR^L$ to obtain a clustering of the T-F bins, such as in deep clustering \cite{hershey2016dpcl}. 
In this work, we are interested in replacing these Euclidean embeddings and the way they are used to derive masks by equivalents in a hyperbolic space, as we hope that this leads to better embeddings that encompass hierarchical relationships between sounds. As a starting point, we focus here on the case of class-based separation with a softmax output layer, as this has been the primary non-linearity used in hyperbolic neural networks in computer vision~\cite{Khrulkov2020CVPR, suris2021learning, Atigh_2022_CVPR}, where the Euclidean embeddings and the multinomial logistic regression (MLR) classification approach applied to them are replaced by hyperbolic embeddings and a hyperbolic MLR. 
We leave the application to other non-linearities and to deep clustering to future work.

\vspace{-.1cm}
\subsection{Hyperbolic embedding-based source separation}

We first briefly review several notions pertaining to Riemannian manifolds and hyperbolic spaces that are useful in introducing hyperbolic embeddings and MLR, following \cite{ganea2018hyperbolicnn,peng2021survey,shimizu2021hyperbolic}.

A Riemannian manifold is defined as a pair consisting of a manifold $\mathcal{M}$ and a Riemannian metric $g$, where $g=(g_x)_{x \in \mathcal{M}}$ defines the local geometry $g_x$ (i.e., an inner product) in the tangent space $T_{x}\mathcal{M}$ at each point $x\in\mathcal{M}$. While $g$ defines the geometry locally on $\mathcal{M}$, it also defines the global shortest path, or \emph{geodesic} (analogous to a straight line in Euclidean space), between two given points on $\mathcal{M}$. %
One can define an \emph{exponential map} $\text{exp}_x$ which projects any vector $v$ of the tangent space $T_{x}\mathcal{M}$ onto $\mathcal{M}$, such that $\text{exp}_x(v) \in \mathcal{M}$, and inversely a logarithmic map which projects any point in $\mathcal{M}$ back onto the tangent space at $x$.

The $L$-dimensional hyperbolic space is an $L$-dimensional Riemannian manifold of constant negative curvature $-c$. It can be described using several isometric models, among which we focus here on the Poincar\'e unit ball model $(\mathbb{D}_c^L,g^{\mathbb{D}}_c)$,  defined in the space $\mathbb{D}_c^L = \{x \in \mathbb{R}^L \ | \ c||x||^2 < 1\} \}$. We assume $c > 0$, such that $\mathbb{D}_c^L$ corresponds to a ball of radius $\frac{1}{\sqrt{c}}$ in Euclidean space.
Its Riemannian metric is given by $g^\mathbb{D}_c(x) = (\lambda^c_x)^2g^E$, where $\lambda^c_x=2/(1-c\|x\|^2)$ is a so-called \emph{conformal factor} and $g^E$ the Euclidean metric. %
Given two points $x,y \in \mathbb{D}_c^L$, their induced hyperbolic distance $d_c$ is obtained as
\begin{equation}
    d_c(x,y) =\frac{2}{\sqrt{c}} \text{tanh}^{-1} (\sqrt{c}\|-x \oplus_c y\|),
\label{eq:hyper_dist}
\end{equation}
where $\oplus_c$ denotes the M\"obius addition in $\mathbb{D}_c^L$, defined as
\begin{equation}
    x\oplus_c y = \frac{(1+2c\langle x,y \rangle +c\|y\|^2) x + (1-c\|x\|^2)y }
    {1+2c\langle x,y \rangle +c^2\|x\|^2\|y\|^2}.
\end{equation}

One way to go back and forth between the Euclidean space $\RR^L$ and the hyperbolic space $\mathbb{D}_c^L$  is to use the exponential and logarithmic maps at the origin $0$, as $T_{0}\mathbb{D}_c^L=\RR^L$, which can be obtained for $v \in \RR^L \setminus \{0\}$ and $y \in \mathbb{D}_c^L \setminus \{0\}$ as:
\begin{equation}
    \exp^c_0(v) = \frac{\text{tanh}(\sqrt{c}\|v\|)}{\sqrt{c}\|v\|}v, \quad \log^c_0(y) = \frac{\text{tanh}^{-1}(\sqrt{c}\|y\|)}{\sqrt{c}\|y\|}y.
\end{equation}

We can thus obtain hyperbolic embeddings $z^h_{t,f} \in \mathbb{D}_c^L$ at the output of a classical neural network such as $f_\theta$ by simply projecting the usual Euclidean embeddings in that way:
\begin{equation}
    z^h_{t,f} = \exp_0(z^e_{t,f})=\exp_0(f_\theta(X)_{t,f}).
\end{equation}
To define a hyperbolic softmax based on these hyperbolic embeddings, the Euclidean MLR can be generalized to the Poincar\'e ball as in \cite{ganea2018hyperbolicnn}. In the Euclidean space, MLR is performed by considering the logits obtained by calculating 
the distance of an input's embedding $z\in\RR^L$ (such as $z=z^e_{t,f}$) to each of $K$ class hyperplanes, where the $k$-th class hyperplane is determined by a normal vector $a_k \in \mathbb{R}^L$ and a point $p_k \in \mathbb{R}^L$ on that hyperplane. Analogously, one can define a Poincar\'e hyperplane $\text{H}^c_{p_k,a_k}$ by considering the union of all geodesics passing by a point $p_k$ and orthogonal to a normal vector $a_k$ in the tangent space $T_{p_k}\mathbb{D}_c^L$ at $p_k$. Hyperbolic MLR can then be defined by considering the distance from a hyperbolic embedding $z=z^d_{t,f}\in \mathbb{D}_c^L$ to each $\text{H}^c_{p_k,a_k}$, leading to the following formulation as shown in \cite{ganea2018hyperbolicnn}: %
\begin{equation}
    p(\kappa\!=\!k|z)\! \propto \exp  \Bigl( \frac{\lambda^c_{p_k}\|a_k\|}{\sqrt{c}} \text{sinh}^{-1} \!
    \Bigl( \frac{2\sqrt{c}|\langle -p_k\!\oplus_c\! z,a_k \rangle|}{(1\!-\!c\|\!-\!p_k \!\oplus_c\! z\|^2)\|a_k\|} \Bigr) \Bigr).
\label{eq:hyp_mlr}
\end{equation}
We can use the probability $p(\kappa=k|z^d_{t,f})$ that T-F bin $(t,f)$ is dominated by the $k$-th source to obtain $K$ source-specific mask values for each T-F bin in the input spectrogram $X$. This procedure is illustrated in Fig.~\ref{fig:overall_example}.
Note that $p_k$ and $a_k$ both parameterize the $k$-th hyperbolic hyperplane and are trainable.
All parameters of the network $f_\theta$ and of the hyperplanes can be optimized using classical source separation objective functions, either on the masks or on the reconstructed signals.

\vspace{-.1cm}
\subsection{Uncertainty in Audio Source Separation}

It has been shown repeatedly \cite{Atigh_2022_CVPR,peng2021survey,ganea2018hyperbolicnn,Khrulkov2020CVPR} that the distance $d_c(0,z^h_{t,f})$ of the projected embeddings to the center of the Poincar\'e ball can serve as a reliable measure of classification certainty. This hyperbolic distance can be computed from the $L_2$ norm of $z^h_{t,f}$ as considered in the Euclidean space using the monotonous relationship $\log((1+c\|z^h_{t,f}\|^2)/(1-c\|z^h_{t,f}\|^2))$, so that the $L_2$ norm can equivalently be considered as a measure of certainty. 
We here aim at validating this hypothesis with audio embeddings and bringing further light onto the notion of certainty in the context of audio source separation. While hyperbolic uncertainty has previously been exploited towards the network optimization stage \cite{jieun2022speechpoincare}, the direct impact of this notion on quality criteria such as amount of artifacts, distortion, or interferences present in resulting signal has yet to be explored.

Furthermore, following the observation presented in \cite{Atigh_2022_CVPR}, we argue that the notion of certainty observed on audio embeddings provides meaningful insights of signal content, such as the complexity of the scene or the number of active sources it may contain. 

\section{Experimental Setup}
\label{sec:exp_setup}
\noindent{\bf Slakh+Speech Dataset}: To test our proposed method we built a simple hierarchical source separation dataset containing mixtures from two ``parent’’ classes - \emph{music} and \emph{speech}, and five ``leaf’’ classes - \emph{bass}, \emph{drums}, \emph{guitar}, \emph{speech-male}, and \emph{speech-female}.  As building blocks, we use the clean subset of LibriSpeech \cite{librispeech_dataset}, and Slakh2100 \cite{manilow2019cutting}, which is a dataset of 2,100 synthetic musical mixtures, each containing  bass, drums, and guitar stems in addition to various other instruments. We built a dataset consisting of 1947 mixtures, each 60~s in length, for a total of about 32 hours. The data splits are 70\%, 20\%, and 10\% for training, validation, and testing sets, respectively.  The \emph{speech-male} source target consists of male speech utterances randomly picked (without replacement) and concatenated consecutively (without overlap) until the 60~s track length is reached. Any signal from the last concatenated utterance exceeding that length was discarded. We used the same procedure for \emph{speech-female}. For Slakh2100, we only selected the first 60~s of the \emph{bass}, \emph{drums}, and \emph{guitar} stems for each track. Any tracks with a duration less than 60~s were discarded. All sources were summed without applying additional gains to make the overall mixture along with the \emph{speech} and \emph{music} submixes. This lead to challenging input SDR values (``No Proc.'' in Table~\ref{table:results_sdr}), with standard deviation values ranging from 2-13 dB depending on the class.

\noindent{\bf Network architecture and training setup}:
Our model consists of four BLSTM layers with 600 units in each direction, followed by a dense layer to obtain an $L$-dimensional Euclidean embedding for each T-F bin. A dropout of $0.3$ is applied on the output of each BLSTM layer, except the last. 
For the hyperbolic models ($c\!>\!0)$, an exponential projection layer is placed after the dense layer, mapping the Euclidean embeddings onto the Poincar\'e ball with curvature $-c$.  As discussed in Section~\ref{sec:hss}, MLR layers, either Euclidean or hyperbolic with softmax activation functions are then used to obtain masks for each of the source classes. In practice, we follow the hierarchical softmax approach from~\cite{Atigh_2022_CVPR}, and have two MLR layers: one with $K\!=\!2$ for the parent (speech/music) sources, and a second with $K\!=\!5$ for the leaf classes.  We use the mixture phase for resynthesis and compare multiple training objectives in Section~\ref{sec:experimental}.

We use the ADAM optimizer for Euclidean parameters, and the Riemannian ADAM \cite{becigneul2018riemannian} implementation from \emph{geoopt} \cite{kochurov2020geoopt} for hyperbolic parameters. All models are trained using chunks of 3.2 s and a batch size of 10 for 300 epochs using an initial learning rate of $10^{-3}$, which is halved if the validation loss does not improve for 10 epochs. We use an STFT size of 32 ms with 50\% overlap and square-root Hann window.

\section{Experimental Analysis}
\label{sec:experimental}
\begin{table}[t]
\scriptsize
\centering
\sisetup{
detect-weight, %
mode=text, %
tight-spacing=true,
round-mode=places,
round-precision=1,
table-format=3.1,
table-number-alignment=center
}
\caption{SI-SDR in dB on the Slakh+Speech test set with different loss functions and embedding dimensions $L$. The ``Hyp.'' column denotes whether the model was trained using hyperbolic ($c=1$) or Euclidean embeddings.}\vspace{-.3cm}
\setlength{\tabcolsep}{2.5pt}
\resizebox{.99\linewidth}{!}{%
\begin{tabular}[t]{l S S S S S S S S S |S}
\toprule
& & & \multicolumn{2}{c}{Parents} & \multicolumn{3}{c}{Music Leaves} & \multicolumn{2}{c|}{Speech Leaves} & \multicolumn{1}{c}{} \\
\cmidrule(lr){4-5} \cmidrule(lr){6-8} \cmidrule(lr){9-10} 
Loss & {$L$} & {Hyp.} & {Music} & {Speech} & {Bass} & {Drums} & {Guitar} & \multicolumn{1}{r}{Male} & \multicolumn{1}{r}{Female} & \multicolumn{1}{c}{Avg.} \\
\midrule
No Proc. & & & -2.94 & 2.99 & -8.84 & -11.8 & -8.73 & -3.17 & -4.16 & -5.24   \\

\midrule
Oracle$_\text{PSF}$ & & & 10.25 &	 13.80 &	5.56 &	 9.79 &	6.98 &	10.10 &	 11.05 &	9.65\\

\midrule
PSA & {2} & $\times$ & 7.73 & 11.28 & 2.70 & 4.49 & \bfseries 2.98 & 6.15 & 6.48 & 5.97\\
WA & {2} & $\times$ & 7.08 & 10.83 & \bfseries 2.80 & 3.18 & 2.33 & 5.49 & 5.64 & 5.34 \\
CE$_\text{IBM}$ & {2} & $\times$ & 5.06 & 8.76 & -2.29 & 1.84 & -1.03 & 2.76 & 2.93 & 2.57  \\
CE$_\text{IBM, W.}$ & {2} & $\times$ & \bfseries 8.03 & \bfseries 11.47 & 2.65 & \bfseries 4.86 & 2.48 & \bfseries 6.44 & \bfseries 6.64 & \bfseries 6.08\\

\midrule
PSA & {2} & \checkmark & 7.46 & 10.87 & -4.34 & -6.88 & -4.61 & 5.97 & 6.28 & 2.11\\
WA & {2} & \checkmark & 7.26 & 10.96 & 2.50 & 3.72 & 2.32 & 5.34 & 5.43 & 5.36 \\
CE$_\text{IBM}$ & {2} & \checkmark & 5.42 & 9.04 & -0.26 & 2.56 & 0.78 & 3.40 & 3.83 & 3.54  \\
CE$_\text{IBM, W.}$ & {2} & \checkmark & \bfseries 7.90 & \bfseries 11.36 & \bfseries 2.93 & \bfseries 5.10 & \bfseries 3.19 & \bfseries 6.52 & \bfseries 6.99 & \bfseries 6.29\\

\midrule
PSA & {128} & $\times$ & 7.73 & 11.16 & \bfseries 3.25 & 4.95 & 3.28 & 6.10 & 6.37 & 6.12\\
WA & {128} & $\times$ & 7.40 & 11.20 & \bfseries 3.30 & 4.80 & 3.01 & 5.89 & 6.11 & 5.96 \\
CE$_\text{IBM}$ & {128} & $\times$ & 6.54 & 10.09 & 1.73 & 6.06 & 2.48 & 5.07 & 5.72 & 5.39  \\
CE$_\text{IBM, W.}$ & {128} & $\times$ & \bfseries 7.93 & \bfseries 11.47 & \bfseries 3.31 & \bfseries 6.84 & \bfseries 3.83 & \bfseries 6.78 & \bfseries 7.30 & \bfseries 6.78\\

\midrule
PSA & {128} &  \checkmark & 7.45 & 10.89 & 2.89 & 4.88 & 3.13 & 5.78 & 5.99 & 5.86\\
WA & {128} &  \checkmark & 6.97 & 11.23 & \bfseries 3.26 & 4.76 & 2.87 & 5.87 & 5.99 & 5.85 \\
CE$_\text{IBM}$ & {128} &  \checkmark & 6.02 & 9.72 & 1.27 & 5.78 & 2.01 & 4.50 & 5.18 & 4.93  \\
CE$_\text{IBM, W.}$ & {128} &  \checkmark & \bfseries 7.80 & \bfseries 11.34 & \bfseries 3.28 & \bfseries 6.13 & \bfseries 3.72 & \bfseries 6.66 & \bfseries 7.09 & \bfseries 6.57\\

\bottomrule
\end{tabular}%
\label{table:results_sdr}
}\vspace{-.4cm}
\end{table}

\noindent{\bf Model comparisons:} Table \ref{table:results_sdr} presents the scale-invariant signal-to-distortion ratio (SI-SDR)~\cite{leroux2019sdr} on the Slakh+Speech dataset described in Sec.~\ref{sec:exp_setup}. We include the no processing condition (lower bound, using the mixture as estimate) and oracle phase sensitive mask~\cite{erdogan2015mask} (upper bound). For loss functions, we use the phase-sensitive approximation (PSA) \cite{erdogan2015mask} with $L_1$ loss \cite{wang2018objective}, and the waveform approximation (WA) loss training through the iSTFT \cite{Wang2018Interspeech09}. Additionally, to compare with related image segmentation approaches~\cite{Atigh_2022_CVPR}, we explore using a cross entropy (CE) loss with the ideal binary mask (IBM) as a training target~\cite{wang2018supervised}, denoted by CE$_{\text{IBM}}$ in Table \ref{table:results_sdr}, and a magnitude-ratio-weighted CE loss inspired by the deep clustering weighting function from~\cite{wang2018objective}, denoted by CE$_{\text{IBM, W.}}$, where the weight for each T-F bin is set to the ratio of the magnitude of the mixture at that bin to the sum over all bins. %

\begin{figure}
    \centering
        \includegraphics[width=1.0\linewidth]{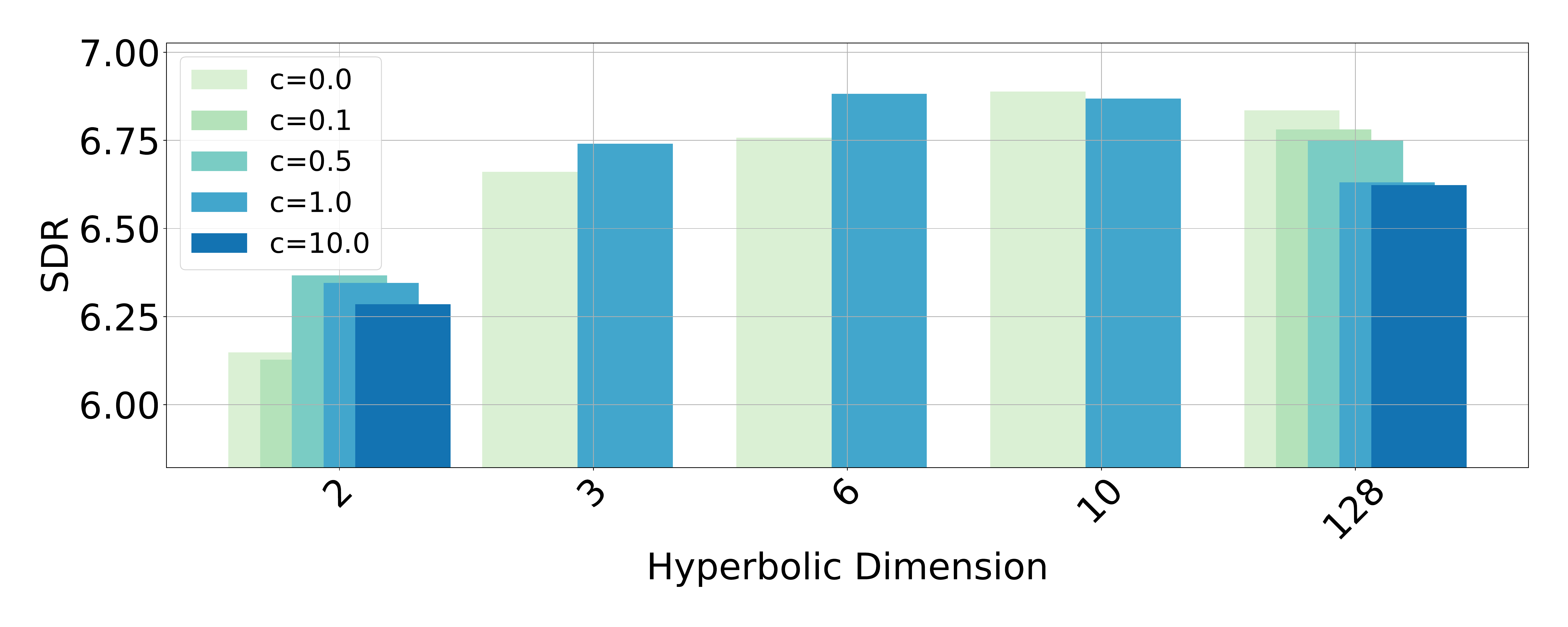}\vspace{-.4cm}
    \caption{SI-SDR [dB] averaged across all (parent and leaf) sources for various curvature values and embedding dimensions. %
    }\vspace{-.3cm}
    \label{fig:curv_comp}
\end{figure}

In Table~\ref{table:results_sdr}, CE$_{\text{IBM}}$ consistently performs poorly, likely because low-energy T-F bins are weighted equally with high energy bins. However, CE$_{\text{IBM, W.}}$ consistently outperforms PSA and WA, which we hypothesize is because the CE objective is particularly well-matched to the softmax mask nonlinearity we use in this work. We also note that, consistent with previous hyperbolic work~\cite{nickel2017pc,ganea2018hyperbolicnn,Atigh_2022_CVPR, jieun2022speechpoincare}, performance of hyperbolic models is relatively stronger than Euclidean ones at low ($L=2$) embedding dimension. This result is further confirmed in Fig.~\ref{fig:curv_comp}, where we explore multiple embedding dimensions and curvature parameters. Unless otherwise stated, the hyperbolic configuration used in all subsequent experiments uses the CE$_{\text{IBM, W.}}$ loss, curvature parameter $c=0.1$, and embedding dimension $L=2$.

\begin{figure}
    \centering
        \includegraphics[width=1.0\linewidth]{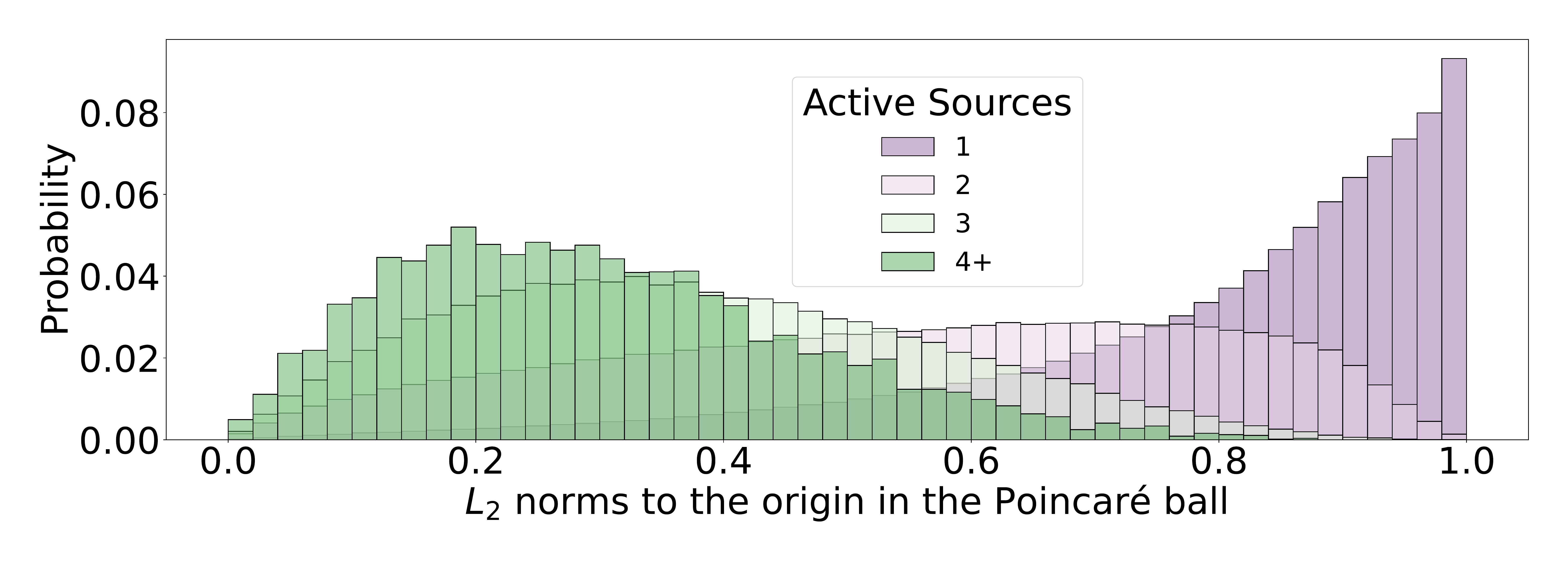}\vspace{-.4cm}
    \caption{Histograms denoting the distribution of $L_2$ norms to the origin of the Poincar\'e Ball for all embeddings in our test-set. We define the number of active sources for each embedding based on the ground-truth mask values of their associated T-F bin.}\vspace{-.5cm}
    \label{fig:semantics}
\end{figure}

\begin{figure}[t]
    \centering
        \includegraphics[width=1.0\linewidth]{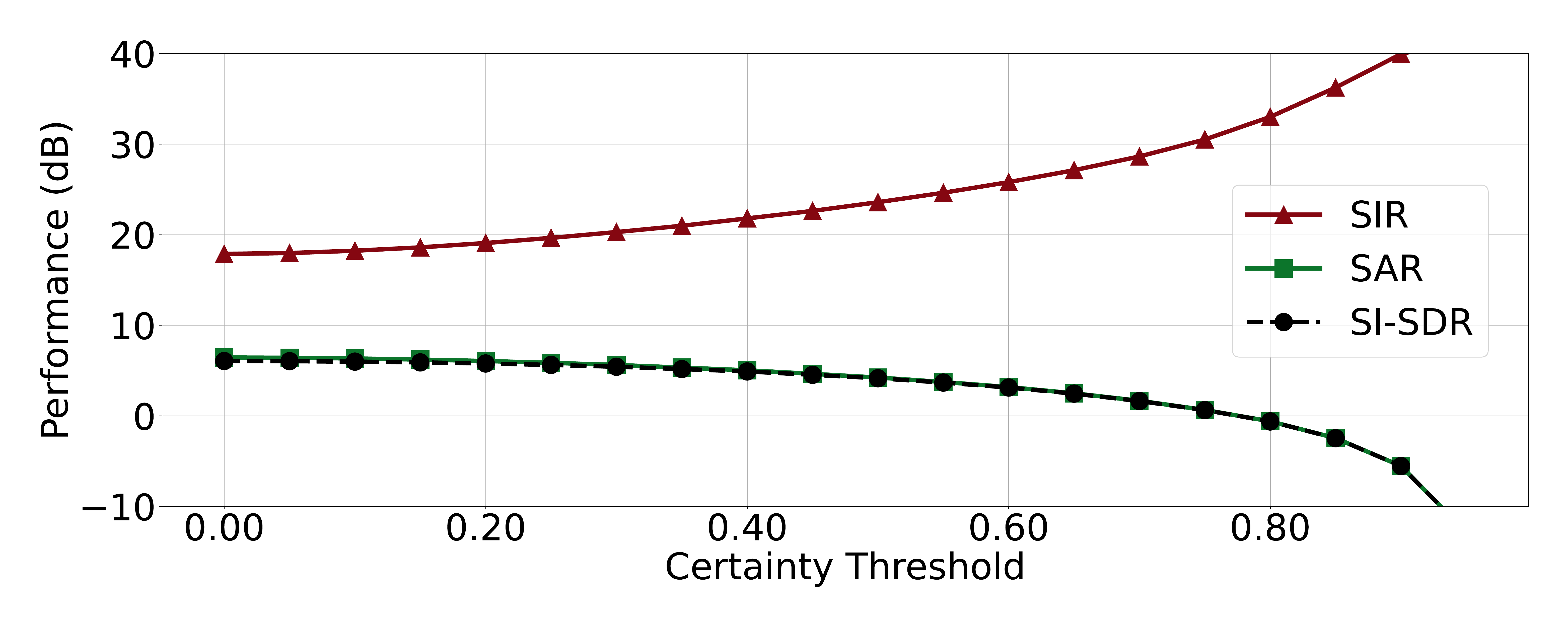}\vspace{-.4cm}
    \caption{Separation metrics as a function of certainty thresholding (here applied to the $L_2$ norm $\|z^h_{t,f}\|_2$). A threshold of $0$ means all hyperbolic embeddings are taken into account during inference, and a threshold of $0.95$ means only embeddings with high certainty (i.e., high distance from the origin) are considered, while bins below the threshold are silenced.}
    \vspace{-.4cm}
    \label{fig:cert_thresh}
\end{figure}

\noindent{\bf Hyperbolic Certainty and Audio Semantics}: In Fig.~\ref{fig:semantics}, we show that embeddings associated with T-F bins containing many active sources (i.e., 4+) tend to be positioned close to the center of the Poincar\'e Ball, where the certainty is low. In contrast, embeddings with a single active source tend to be located closer to the edge of the Poincar\'e Ball, where the certainty is high. Intuitively, the higher the number of sound classes (i.e., interferences) active, the more uncertain as for what specific single source the bin embedding may belong to. We consider a source active for a given T-F bin if it satisfies the following conditions: (1) it is non-silent, i.e., it is not more than $20.0$ dB below the maximum T-F bin energy in a given sound file, and (2) the ratio of the ground-truth T-F bin magnitude for that source divided by the sum of ground-truth magnitudes for all sources (i.e., the ideal ratio mask) is greater than $0.1$.

\noindent{\bf Hyperbolic Distance - Interference and Artifacts}: We also explore a simple approach for using hyperbolic certainty estimates to modify and interact with mask values. Specifically, we embed all T-F bins for a given mixture in the Poincar\'e ball, and any bins that are close to the origin (i.e., low certainty) have their mask values set to zero.
Figure~\ref{fig:cert_thresh} shows the impact of varying this certainty threshold on the SI-SDR, SIR, SAR~\cite{leroux2019sdr,vincent2006bsseval} metrics. Each data point denotes an evaluation pass on the entire test set given a certainty threshold (x-axis). We see that if we silence uncertain T-F bins in the mask predictions, the resulting signals will contain less interference at the expense of an increase in artifacts. Using hyperbolic certainty to control the trade-off between artifact introduction and interference reduction could be an exciting area to further explore in future work.

\noindent{\bf Comparison with Bayesian certainty:} To further validate our hyperbolic certainty measurements, we lead an experiment comparing hyperbolic and Bayesian certainty. The latter approach is generally achieved by means of Monte-Carlo dropout \cite{yarin2015mcdropout,nazreen2018mcdropout} where multiple stochastic forward passes using a trained network with different dropout realizations are performed to obtain the posterior distribution for a single input example. The network prediction and its associated certainty map can then be inferred from the mean and variance computed over all stochastic passes. In order to deal with the fact that our network makes predictions over multiple sound classes, we follow the approach presented in \cite{Mukhoti2018EvaluatingBD} and formulate the Bayesian certainty $\zeta_{t,f}$ at T-F bin $(t,f)$ as the negative predictive entropy
\begin{equation}
\label{eq:bayesian}
\zeta_{t,f} \!=\!\! \sum_{k=1}^{K_{\text{leaf}}} \Bigl(\frac{1}{N}\!\sum_n p(\kappa\!=\!k|\hat{z}^{(n)}_{t,f})\Bigr)  \text{log} \Bigl(\frac{1}{N}\!\sum_n p(\kappa\!=\!k|\hat{z}^{(n)}_{t,f})\Bigr),\!\!
\end{equation}
where $K_{\text{leaf}}$ denote the number of leaf classes (i.e., without the parent classes), $N$ the total number of stochastic passes, and $p(\kappa=k|\hat{z}^{(n)}_{t,f})$ the probability, expressed as the softmax output, that T-F bin $(t,f)$ belongs to source class $k$, and $\hat{z}^{(n)}_{t,f}$ is the embedding obtained with the dropped out network parameters for stochastic pass $n$. Following \cite{Mukhoti2018EvaluatingBD}, we set the dropout rate to $0.5$ at the output of each BLSTM layer of our network. A low entropy (i.e., high certainty) means that the predictability is high, while a high entropy (low certainty) means that mask values fluctuate significantly across passes.

\begin{figure}
    \centering
        \includegraphics[width=1.0\linewidth]{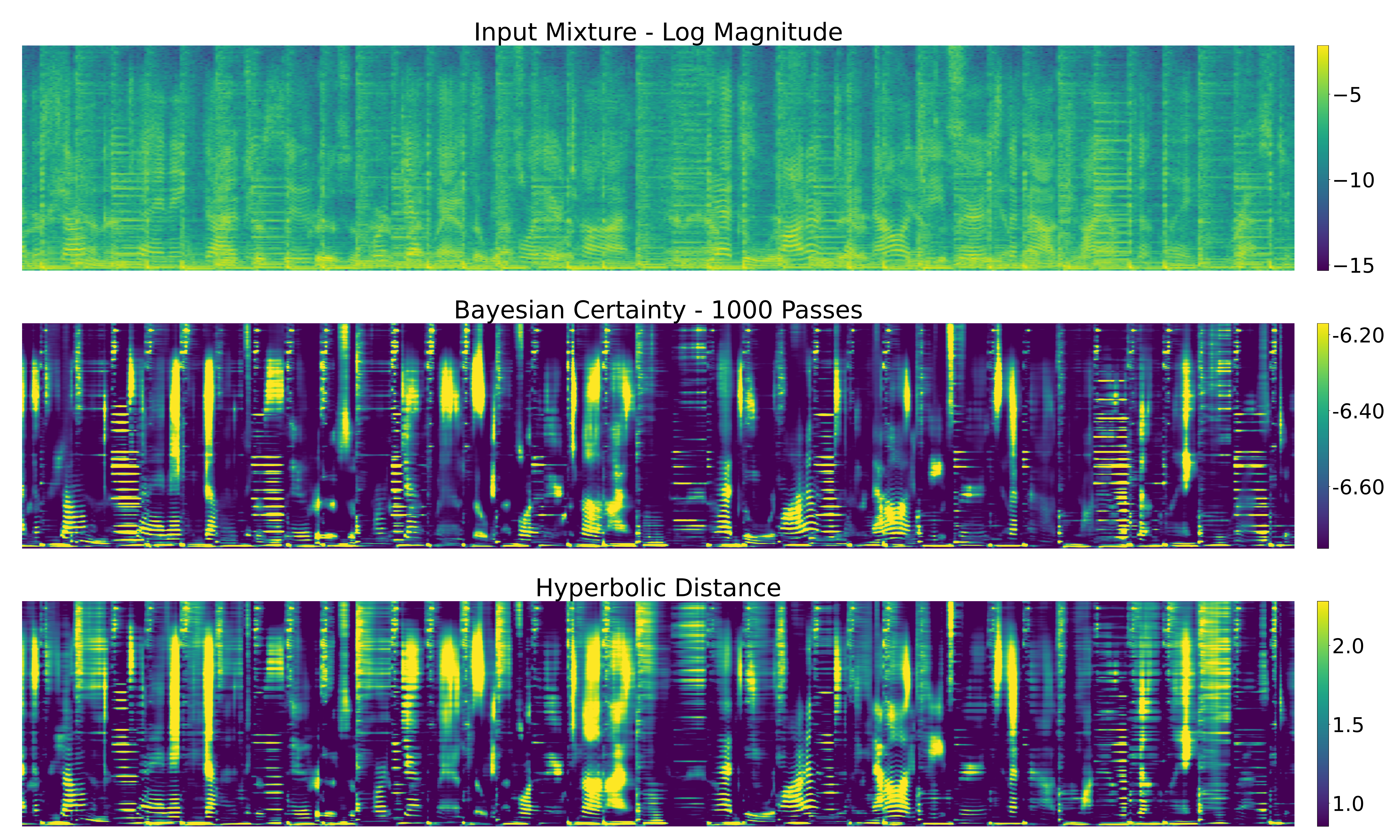}\vspace{-.1cm}
    \caption{Comparison between Bayesian and hyperbolic certainty. The Bayesian certainty is computed by running 1000 passes with Monte-Carlo dropout during inference and computing the bin-wise negative entropy across all passes on the five leaf nodes (i.e., excluding the parent nodes). The hyperbolic certainty is obtained by computing the hyperbolic distance from the Poincar\'e ball origin $d_c(0,z^h_{t,f})$ for each T-F bin embedding $z^h_{t,f}$.}\vspace{-.3cm}
    \label{fig:bayesian}
\end{figure}

Fig. \ref{fig:bayesian} provides a visual contrast of the certainty map computed using 1000 Monte-Carlo dropout iterations and the one obtained using the hyperbolic distance from the Poincar\'e ball origin for each T-F embedding in a single forward pass. The colorscale limits on both maps are set using the 30th and 95th percentiles to ensure a fair qualitative comparison, even though the scales are different. We observe a clear resemblance %
between both maps. The correlation coefficient between the two certainty maps is $\rho=0.75$. This observation tells us that the hyperbolic certainty map, which comes for free (i.e., $N=1$), can be as interpretable as its Bayesian counterpart, which requires multiple forward passes at inference time. 

A demo of a hyperbolic separation interface is available~\footnote{\url{https://darius522.github.io/hyperbolic-audio-sep/}}.

\section{Conclusion}
\label{sec:conclusion}
We have investigated the use of the Poincar\'e Ball model to perform audio source separation in the hyperbolic space. Our hyperbolic model operates and computes T-F embeddings in the Euclidean space and projects them onto the hyperbolic space. Masks are obtained by hyperbolic multinomial logistic regression considering the distance from hyperbolic embeddings to hyperbolic hyperplanes.
Through our experimental setup, we have demonstrated that hyperbolic audio embeddings can convey useful information in regards to uncertainty and underlying hierarchical sound structures. We associated these notions to known audio concepts such as artifacts and interferences.
In the future, we aim at exploring how the hyperbolic space, and especially the notion of hierarchy, could benefit additional audio-related tasks, such as audio tagging and sound-event detection. In the context of audio source separation, we believe that further connections to deep clustering, and more complex and deeper sound taxonomies are worth exploring.

\bibliographystyle{IEEEtran}
\bibliography{refs}

\end{document}